\documentclass[amsmath,amssymb,,showpacs,twocolumn]{revtex4}
\usepackage{color}
\usepackage{bm}
\usepackage{epsfig}
\usepackage{amsmath}
\usepackage{amsfonts}
\usepackage{amssymb}
\usepackage{bezier}

\def\beq{\begin{equation}}
\def\eeq{\end{equation}}
\def\be{\begin{equation}}
\def\ee{\end{equation}}
\def\bea{\begin{eqnarray}}
\def\eea{\end{eqnarray}}
\def\half{\mbox{$1\over2$}}

\def\bA{{\bf A}}

\def\bJ{{ J}}

\def\bk{{\bf k}}

\def\bx{{\bf x}}

\def\br{{\bf r}}

\def\bq{{\bf q}}

\def\hO{\hat{O}}

\def\cO{{\cal O}}

\def\cL{{\cal L}}
\def\cH{{ H}}

\def\cL{{\cal L}}

\def\cG{{\cal G}}

\def\hbx{{\hat{\bx}}}

\def\hx{{\hat{x}}}

\def\sd0{\rho^0_s(T)}

\def\1{\mbox{1\hskip-.25em l}}
\def\6{\langle }
\def\9{\rangle }

\def\tr{\textrm{Tr}}
\def\im{\textrm{Im}}
\def\re{\textrm{Re}}

\def\ze{z_{\rm eff}}

\begin{document}
\parindent=0pt

\title{``Bad Metal'' Conductivity of  Hard Core Bosons}
\author{Netanel H. Lindner and Assa Auerbach}
\affiliation{Physics Department, Technion, 32000 Haifa,
Israel}
\begin{abstract}
Two dimensional hard core bosons suffer strong scattering in the
high temperature resistive state at half filling. The dynamical
conductivity $\sigma(\omega)$ is calculated using  non
perturbative tools such as  continued fractions,  series
expansions and exact diagonalization. We find a large temperature range with  linearly increasing resistivity
and broad dynamical conductivity, signaling a breakdown of  Boltzmann-Drude quasiparticle transport
theory.  At zero temperature,  a high frequency peak in
$\sigma(\omega)$ appears above a {\em ``Higgs mass''} gap, and
corresponds to order parameter magnitude fluctuations. We discuss the apparent
similarity between conductivity of hard core bosons  and  phenomenological
characteristics of cuprates, including the universal scaling of  Homes {\em et al.} (Nature \textbf{430},
539 (2004)).
\end{abstract}
\maketitle

\section{Introduction} \label{Introduction}
Quantum transport in condensed matter is largely based on the paradigms of Fermi and Bose gases.
Boltzmann equation for the conductivity is valid in the weak scattering regime where it yields a Drude form \cite{Ziman, Mahan},
\be
\sigma^{Drude} (T,\omega)= {q^2n\over m^*} \re \frac{\tau}{1-i\omega\tau},
\label{drude}
\ee
where $T$ is  temperature, $ \omega$ is  frequency, and  $q,n, m^*,\tau(T)$ are charge,  density, effective mass and scattering time of the constituent quasiparticles.

Interacting bosons in a  strong periodic potential may suffer
strong enough scattering which invalidates  Boltzmann-Drude
theory. An example  is provided by the two dimensional Hard Core
Bosons (HCB)  model  at half filling. While it is established that
the ground state is a {\em bone fide}  superconductor
\cite{Loh,DM,ding,sandvik,Troyer}, the resistive  (normal) phase  involves strongly
interacting  bosons and vortex pairs. Previous work \cite{LAA,Solvay}  showed that
the lattice (umklapp) scattering dramatically
increases  vortex mobility. At half filling, it  produces an abrupt reversal of the
Hall conductivity, and doublet degeneracies associated with each vortex.

HCB models may be experimentally  relevant to  cold  atoms in optical lattices~\cite{PS, Zoller},
underdoped cuprate superconductors~\cite{Uemura,EK-Uemura,Mihlin}, low capacitance Josephson junction arrays~\cite{Ood,Altman-JJ},  and
disordered superconducting films~\cite{fisher}. An added advantage of HCB, is that they are described by
a quantum spin-half $XY$ model which is amenable to tools of quantum magnetism.

It is the purpose of this paper to compute the  conductivity of HCB at half filling.
 We apply and test a set of non perturbative approaches, including continued fraction representation,  series expansions and exact diagonalization.
By studying the dynamical structure of the Kubo formula, we can construct well-converging approximants which agree with high accuracy
sum rules in a wide regime of temperature.
The conductivity at high temperatures is obtained to  order of $T^{-3}$. Near  the superconducting transition, it is
matched  with the critical conductivity which was derived by  Halperin and Nelson~\cite{HN}.
At zero temperature, the dynamical conductivity is obtained from the relativistic Gross-Pitaevskii field theory,
and a variational fit to  12th order moments.

\begin{figure}
\vspace{-0.3cm}
\begin{center}
\epsfxsize=.55\textwidth \centerline{\epsffile{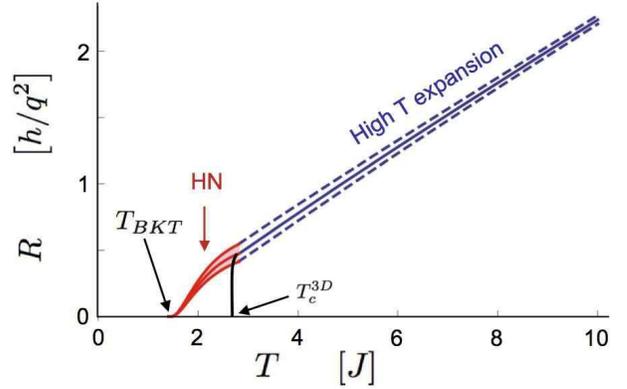}}
\vspace{-0.5cm}
\caption{Temperature dependent resistivity of two dimensional Hard Core Bosons at half filling.
High temperature line (blue online) is calculated up to order $1/T$, with error margins
depicted by  dashed lines. The critical region above the BKT transition, (red online)  is given by the vortex plasma
theory of  Halperin and Nelson (HN). For illustration, a layered system with  weak interlayer coupling shows a rapid rise in resistivity
above the three dimensional transition temperature   $T_c^{3D}$, as depicted by a solid black line.}
\label{fig: resistivity}
\end{center}
\end{figure}


Our key results pertain to the qualitative effects of strong scattering on the conductivity.
 At high temperatures,  the conductivity goes as
\be
\sigma(T,\omega) \approx  0.91 {q^2 \over h   } {\tanh(\hbar\omega/(2T))\over (\omega/\bar{\Omega})} \exp\left( -\left({\omega\over \bar{\Omega}}\right)^2 \right),
\label{sigma-app}
\ee
where $\bar{\Omega}$ is a high frequency scale.
Eq. (\ref{sigma-app}), when fit at low frequencies and temperatures, to the form  (\ref{drude}),
yields an ``effective scattering rate''  which is equal to $2T$.
The resistivity as shown in Fig.~\ref{fig: resistivity} exhibits
 ``bad metal'' ~\cite{badmetal} behavior: it exceeds the boson Ioffe-Regel limit~\cite{res-sat2} of $R>h/q^2$, without saturation~\cite{res-sat1,res-sat11}.
The resistivity's linear slope defines the proportionality coefficient between the zero temperature superfluid stiffness,
 and product of $T_c$ and the normal state conductivity near $T_c$.  We note that  linearly increasing resistivity~\cite{linear} and an analogous scaling
 of superfluid stiffness with conductivity
 called ``Homes law''~\cite{homes},  have been observed in the  cuprate family, as depicted in Fig.~\ref{fig:homes}.

 In addition, at zero temperature we find a small conductivity peak  above a ``Higgs mass'' gap.
This peak is associated with  order parameter magnitiude fluctuations. These are analogous to coherence oscillations
observed in cold atoms during rapid  Mott to superfluid quenches ~\cite{OSF1,OSF2,OSF-Theory}.
We speculate that perhaps the {\em mid infra-red}  peak observed in some cuprates at low temperatures~\cite{MIR},
might arise from these magnitude  fluctuations.

However we emphasize that similarities between HCB and cuprates  are only suggestive. This work
{\em does not} include  effects of fermion quasiparticles,  magnetic excitations, and inhomogeneities
which are experimentally important.

This paper is organized as follows: Section \ref{sec:HCB} introduces the  HCB model and  lists some of the established thermodynamics.
Section  \ref{sec:Kubo1} discusses the Kubo formula in general, and  derives the moment expansion, continued fraction recurrents, and orthogonal polynomials
which can be used to evaluate $\sigma$ non perturbatively and generate a high temperature expansion. Section \ref{sec:Kubo2} derives the
particular recurrents for the HCB at half filling. Justification for the rapidly converging  harmonic oscillator expansion is provided.
 Sections  \ref{sec:Kubo1},\ref{sec:Kubo2}  are quite technical, and could be avoided at first reading.
The results of our calculations are provided in the following sections.
Section \ref{sec:DCHT} plots the dynamical conductivity at high temperatures, and explains three approaches which converge to the same curve.
 Section \ref{sec:DC} obtains the  resistivity as a function of temperature.
 Section \ref{sec:T=0} obtains the dynamical conductivity and ``Higgs mass''  at zero temperature. Section \ref{sec: discussion} summarizes
the key results and discusses their possible relevance to the dynamical and DC  conductivity of cuprate superconductors.

\section{Hard Core Bosons}
 \label{sec:HCB}
 Hard core bosons  are defined on lattice sites $i=1,\ldots N$
with restricted occupation numbers, $n_i=0,1$. The  constrained creation operators  $\tilde{a}_i^\dagger$ are  represented by spin half raising operator
\bea
\tilde{a}_i^\dagger &=&  S_i^+=S_i^x+iS_i^y,\nonumber\\
\tilde{a}_i &=&  S^-_i.
\eea
Thus, their commutation relations are
\bea
\left[ S_i^-, S_j^+\right]  &=& -   \delta_{ij} 2 S^z_i\nonumber\\
\leadsto ~  \left[ \tilde{a}_i, \tilde{a}^\dagger_j\right] &\equiv &  \delta_{ij} \left( 1-2n_i \right).
 \label{HCB-CR}
 \eea

A minimal model of  HCB hopping with Josephson coupling $J$,  coupled to an electromagnetic field $A_{ij}$,
is  the gauged $S=\half$ quantum $XY$ model,
\bea
 \cH &=& -2J \sum_{\6 i,j \9}
\left(e^{iqA_{ij} } \tilde{a}_i^\dagger \tilde{a}_j  +{\rm H.c.}\right)\nonumber\\
&=& -2J \sum_{\6 i,j \9}
\left(e^{iqA_{ij} } S_{i}^{+}S^{-}_{j} +{\rm H.c.}\right),\nonumber\\
\label{xxz}
\eea
where $q$ is the boson charge ($=2e$, for electronic  superconductors) and  we use units of $\hbar=c=1$.
 Here we consider
$\6i,j\9$ to be nearest neighbor bonds on the square lattice.
In the absence of a chemical potential (i.e. a Zeeman field coupled to $\sum_i S_i^z$), the Hamiltonian describes a density of half filling
(zero magnetization), with  half a boson per site.

The uniform current operator, for $\bA=0$,  is given by
\bea
\bJ_x  &=&  -{ 2 i q J\over\sqrt{N}} \sum_{\6 i,j \9}\left(  \tilde{a}_i^\dagger \tilde{a}_j  - \tilde{a}_j^\dagger \tilde{a}_i \right) \nonumber\\
&=&  { 4 q J \over \sqrt{N}} \sum_\br \left(S^x_\br S^y_{\br+\hbx}-S^y_\br S^x_{\br+\hbx}\right).
\label{jx}
\eea

We note that in one dimension, the current of HCB  is a conserved operator since,
\be
\left[ H^{1D}, J_x \right] =0,
\ee
and hence the real conductivity trivially vanishes at all finite frequencies and temperatures.
In two and higher  dimensions, this is not the case: the conductivity has non trivial dynamical structure.

Below we review some established results for the thermodynamic  properties of the two dimensional  quantum XY model which are relevant to
the conductivity.

\subsection{Superfluid stiffness}
It is widely believed that at zero temperature $H$ has long range order.
Thus, at low temperatures $T\ge 0$,  the
{\em two dimensional boson superfluid stiffness $\rho_s$}, which has units of energy,  is finite:
\be
 \rho_s \equiv   q^{-2} {d^2 F(T,n) \over (dA_x)^2}\Bigg|_{\bA=0} >0,
\ee
where $F$ is the free energy in the presence of a uniform   field  $\bA=A_x \hbx$.
A classical  (mean field) approximation at zero temperature yields  a non-monotonous density dependence,
\be
\rho_s^{cl}(0,n)= 4J n(1-n).
\label{rho-cl}
\ee
where $n$ is the mean boson occupation (filling).
Half filling $n=\half$ is {\em ``optimal''}, with maximal $\rho_s$.
Quantum corrections  to $\rho_s^{cl}(0,\half)$ enhance it  by
about 7\%~\cite{sandvik,Troyer}.

\subsection{BKT Transition}
The {\em static} order parameter correlations of (\ref{xxz}) are  described by a renormalized classical $XY$ model.
 At low temperatures, correlations decay as a  power-law in distance.
$\rho_s(T)$ decreases with $T$,  until it falls discontinuously to zero at $T_{BKT}$,
the  Berezinskii-Kosterlitz-Thouless (BKT) transition temperature~\cite{KT, KT2, berez}.
At half filling,  quantum Monte-Carlo (QMC) simulations of (\ref{xxz}) have determined~\cite{DM,ding},
\be
T_{BKT} \simeq 1.41 J.
\ee
Just below the transition, a universal relation holds:
\be
 \rho_s(T_{BKT})  = {2\over \pi} T_{BKT}.
\ee
For $T>T_{BKT}$,  $\rho_s=0$, and the correlation length is
\be
\xi \sim A \exp\left( {B\over \sqrt{ (T-T_{BKT})/T_{BKT} }} \right),
\label{xiT}
\ee
where $A=0.285$ and $B=1.92$  as determined by QMC ~\cite{ding}.

In  multilayered systems  with weak interlayer coupling~\cite{HT,Erica, Mihlin} $J_c<<J$, the 3D  transition temperature
 is higher than $T_{BKT}$ by a factor,
\be
T^{3D}_c = T_{BKT} \left( 1+ {B^2\over \log^2(0.144~J/J_c)} \right).
\label{T3D}
\ee

\subsection{Boson particle-hole symmetry}
The charge conjugation operator
\be
C=e^{i\pi \sum_i S^x_i},
\ee
transforms particles into oppositely charged holes:
\bea
\tilde{a}_i &\to &\tilde{a}^\dagger_i.\nonumber\\
n_i &\to& (1-n_i),\nonumber\\
H[q\bA,n] &\to & H[-q\bA,1-n].
\label{ph}
\eea
It follows therefore that under reflection about half filling,
the Hall conductivity reverses sign, while the superfluid stiffness and longitudinal conductivity are invariant.

In the low density limit  $n<<\half$,  the  HCB are effectively
unconstrained~\cite{Mincas,Fetter} as seen by  (\ref{HCB-CR}). This can  be  demonstrated
by expanding  the Holstein-Primakoff bosons representation of spin half,
\be
\tilde{a}_i = b^\dagger_i  \sqrt{1-n_i^b} \approx   b^\dagger_i   \left(1 -\half  n_i^b +\cO(n_i^b)^2\right).
\label{HP-ex}
\ee
Truncating the expansion (\ref{HP-ex}), and inserting it  into $H$,
turns it into an interacting  {\em soft bosons}
model. At low densities, the low excitations of  $H$ do  not feel the lattice. Thus,
the  long wavelength properties are well described by
the Galiliean invariant Gross-Pitaevskii (GP) field theory~\cite{PS}, and the mean field superfluid stiffness goes as
\be
\rho_s \sim  4J n \equiv  {\hbar^2 \over m_b a^2} n,
\ee
where $m_b,a$ are the boson  effective mass and lattice constant respectively.
Similarly, by particle hole transformation (\ref{ph}),  $H$  simplifies into a model of weakly interacting holes
as $n \to  1$.

It is around half filling, however,
that higher order interactions  in (\ref{HP-ex}) and lattice effects become relevant.
The Galilean invariant (non-relativistic) GP theory fails to account for important umklapp scattering processes which lead to Hall effect cancellation, and doublet degeneracies of vortex states ~\cite{LAA}.

\section{Non Perturbative analysis of the Kubo Formula}
 \label{sec:Kubo1}
 \subsection{Current fluctuations function}
Thermodynamic averages are denoted by
\be
\6 (\cdot) \9_\beta ={1\over Z} \tr
\left[e^{-\beta H} (\cdot) \right], ~~~Z=\tr e^{-\beta H},
\ee
where $\beta=1/T$. The linear response Kubo formula for the
longitudinal  dynamical  conductivity is ~\cite{scalapino}
\beq
\sigma(\beta, \omega)=i \frac{\6- q^2 K_x\9_\beta-\Lambda_{xx}(\beta,\omega+i\epsilon)}{\omega+i\epsilon},
\label{Kubo}
\eeq
where
\be
K_x = -{4J\over N}
 \sum_\br \left( S_{\br}^{x}S^{x}_{\br+\bx} +S_{\br}^{y}S^{y}_{\br+\bx}\right),
 \ee
is the $x$-kinetic energy, and
$\Lambda $ is the  retarded current-current
response function
\bea
\Lambda_{xx}(z)&=& \int_{0}^\infty dt e^{i z
t}\langle\left[\bJ_x(t),\bJ_x(0)\right]\rangle_\beta \nonumber\\
\bJ_x(t) &\equiv&   e^{i\cH t} \bJ_x e^{-i\cH t},
\label{eq: Lambda}
\eea
where  the HCB current operator is given by (\ref{jx}).

The (real) conductivity is given by
\bea
\sigma(\beta,\omega)&=& q^2\pi \rho_s(\beta) \delta(\omega)+{\tanh(\beta\omega/2)\over \omega} G''(\beta,\omega)\nonumber\\
G''(\beta,\omega)&=& \half \int_{-\infty}^\infty dt e^{-i\omega t} \langle\{\bJ_x(t),\bJ_x(0)\}\rangle_\beta .
\label{Gpp}
\eea
where $\{ \cdot ,\cdot \}$ is an anticommutator.

The {\em current fluctuations function}  $G''(\beta,\omega)$ of HCB at half filling,  is our primary object of attention.
We cannot rely on a nearly free quasiparticle basis about which to expand Eqs.~(\ref{xxz}) or  (\ref{jx}).
We therefore turn our attention  to  non perturbative approaches.

\subsection{Moments expansion}
\label{section: HOE}
It is advantageous in our case, to analyse the dynamical structure of $G''(\omega)$ in the Operator Hilbert Space (OHS).
The OHS is a linear space of HCB (spin half) operators, denoted by capital roman letters,  $A,B,\ldots$, which are the ''hyperstates'' of the OHS.
In this paper, we use  two different  inner products.\\
(i) The infinite temperature product,
 \be
\left( A,B \right)_\infty  = {1\over 2^N} \tr \left ( A^\dagger  B \right).
\label{IP}
\ee
(ii) The zero temperature product,
 \be
\left( A,B \right)_0  = \langle 0|  A^\dagger B |0\rangle,
\label{IP0}
\ee
where $|0\rangle$ is the ground state of the Hamiltonian $H$.
It is easy to verify that both definitions obey the Hilbert space conditions for an  inner  product.
Henceforth, we unify the notations,  and  drop the  subscripts  $0,\infty$.

{\em  ``Hyperoperators''}, denoted by capital script letters, are linear operators which act on hyperstates of  the OHS.
The {\em  Liouvillian} hyperoperator $\cL$ is defined by its action on any hyperstate  $A$ as,
\be
\cL  A \equiv   [H,A].
\label{Li}
\ee
By   hermiticity  of $H$, and the cyclic property of the trace, the Liouvillian $\cL$ is   Hermitian for {\em  both} definitions of the inner products (\ref{IP},\ref{IP0}),
\be
(A,\cL B) = (B,\cL A)^*.
\label{Herm}
\ee
Therefore, $\cL$ has a real eigenspectrum.

The time dependent current operator in the Heisenberg representation $J_x(t)$,
is compactly expressed  using the evolution  hyperoperator,
\be
\bJ_x(t)=   e^{i\cL t} ~\bJ_x.
\label{j(t)}
\ee
The  hyper-resolvent $\cG(z)$,
\beq
\cG(z)={1\over z-\cL},
\eeq
is related to the evolution hyper-operator by
\be
e^{i\cL t}  =\oint {dz\over 2\pi i} e^{i z t}\cG(z).
 \ee
The  contour surrounds the spectrum of $\cL$, which by Eq.
(\ref{Herm}) lies on the real axis.

The complexified current fluctuations function is
\beq
G(\beta,z)= \frac{1}{Z} \tr \left(  e^{-\beta H}
\left\{\bJ_x,\cG(z) \bJ_x \right\}\right).
\label{cG}
\eeq
Eq. (\ref{Gpp}) is recovered by its imaginary part on the real axis,
\be
G''(\beta,\omega)= - \frac{1}{Z}\im \tr\left( e^{-\beta H}
\left\{\bJ_x,\frac{1}{\omega-\cL+i\epsilon }\bJ_x\right\}\right).
\label{LR}
\ee

A direct $1/\omega$ expansion of $(\omega-\cL)^{-1}$ does not
yield the required imaginary function~\cite{Betts}.  To extract
$G''$ one uses complex analysis:
\be
\oint dz z^k \cG(z) = \oint {dz\over z} z^k \sum_{n=0}^\infty \left( {\cL\over z}\right)^n =2 \pi i \cL^k,
\ee
and take the contour around the real axis, to obtain the sum rules for all $k=0,2,4\ldots \infty$,
\be
\int_{-\infty}^\infty {d\omega\over \pi} \omega^k  G''(\beta,\omega)=  \langle \left\{ J_x, \cL^k J_x \right\} \rangle_\beta .
\equiv   \mu_k(\beta).
\label{moments}
\ee
All odd-$k$  moments vanish by symmetry of $G''(\omega)$. The moments
$\mu_k$ are static (equal time) correlators, which can be evaluated numerically or by series expansions.
It is possible, in general,  to compute only a  finite number of  $\mu_k$'s.
The remaining task is to derive converging approximants to $G''(\beta,\omega)$ by, in a sense, inverting Eq. (\ref{moments}).

\subsection{The Liouvillian Matrix}
In order to  invert (\ref{moments}),  we use the structure of the  Liouvillian matrix.
A tridiagonal matrix representation of the Liouvillian is constructed as follows.
We define the root hyperstate by the current operator,
\be
\hO_0= {1\over \sqrt{(\bJ_x,\bJ_x)}} ~\bJ_x.
\ee
An orthonormal set of hyperstates $\hO_n$ can be generated
by sequentially  applying $\cL$  and orthonormalizing by the Gram-Schmidt procedure,
\bea
\hO_{n+1} &\equiv &c_{n+1}\left(  \cL \hO_n - \Delta_{n} \hO_{n-1}\right), ~~~n=1,2,\ldots \nonumber\\
\Delta_n &=&  \left( \hO_n, \cL \hO_{n-1} \right), \nonumber\\
c_{n+1} &=& \left( |\Delta_n|^2 + \left(\hO_n ,\cL^2 \hO_n \right)\right)^{-1/2},
\label{GS}
\eea
where $\Delta_0=0$. Since $J_x$ is hermitian, and $\cL J^x$ is antihermitian, all $\hO_n$ can be chosen to be hermitian (antihermitian)
for even (odd ) $n$, since $\Delta_n$ are purely imaginary. Thus it is easy to prove, for both inner products defined in (\ref{IP}) and (\ref{IP0}),
that $\cL$ has no diagonal matrix elements $\left(\hO_n,\cL \hO_n\right)=0$. Also,
it is straightforward to   prove by induction that $|\hO_n\rangle$ is an orthonormal set:
\be
\left(\hO_n, \hO_{n'}\right) = \delta_{nn'}, ~~~~n,n'=0,1,\ldots \infty.
\label{basis}
\ee
In this basis, $\cL$ is given by the {\em Liouvillian matrix},
\beq
L_{nn'} \equiv \langle \cO_n| \cL |\cO_{n'}\rangle  =
\left(\begin{array}{cccc}
0& \Delta_1 &0 &0 \\
\Delta_1^*  & 0& \Delta_2  & 0\\
0 & \Delta_2^*& 0  & \Delta_3 \\
0 & 0 & \Delta_3^* & ...
\end{array}\right)
\label{eq: cLmat}
\eeq
In both $T=0,\infty$ limits,
the current fluctuations function (\ref{cG}) is given by the root expectation value of the resolvent:
\bea
G(z) &=& \left(J_x,  (z-\cL)^{-1} J_x\right) \nonumber\\
&=& \mu_0 \left(z-L\right)^{-1}_{00}.
\label{Gz0}
\eea
 where
$\mu_0  = (J_x,J_x) $ is the corresponding zeroth moment.

In these limits, all moments of Eq. (\ref{moments})  are root  expectation values:
\be
\mu_k = \left( J_x, \cL^k J_x \right).
\ee
Using (\ref{eq: cLmat}), an explicit, and useful relation between recurrents and moments is obtained
\bea
 \mu_{k}[\Delta]  &=&  \mu_0 \left( L^{k} \right)_{00},\nonumber\\
 \mu_2 &=& |\Delta_1|^2,\nonumber\\
\mu_4 &=& |\Delta_1|^4 + |\Delta_1|^2 |\Delta_2|^2, \nonumber\\
\mu_6&=&  |\Delta_1|^6 + 2|\Delta_1|^4 |\Delta_2|^2 +  |\Delta_1|^2 |\Delta_2|^4  \nonumber  \\
&\phantom{=}&  +|\Delta_1|^2 |\Delta_2|^2|\Delta_3|^2,\nonumber\\
\vdots&=&~~~~.
\label{mom-rec2}
\eea
\subsection{Continued fraction representation}
Inverting $z-L$ in Eq. (\ref{Gz0}) using elementary algebra yields the continued fraction representation~\cite{mori,lee}
\beq
G(z)= \mu_0  \frac{1}{z-\frac{|\Delta_1|^2}{z-\frac{|\Delta_2|^2}{z- ...}}}.
\label{eq: cont frac}
\eeq
At zero temperature we obtain,
\be
G_{T=0}(z) =  2  \langle 0|J_x^2|0\rangle\frac{1}{z-\frac{|\Delta^0_1|^2}{z-\frac{|\Delta^0_2|^2}{z- ...}}},
\label{CF0}
\ee
where $|0\rangle$ is the ground state of $H$.
Similarly, in the high temperature limit,  the leading order $G(z)$ is given by
 \be
G_{T=\infty}(z) = {2 \over 2^N} \tr \left( J_x^2\right)\frac{1}{z-\frac{|\Delta^\infty_1|^2}{z-\frac{|\Delta^\infty_2|^2}{z- ...}}}.
\label{CFinfty}
\ee
The infinite list of recurrents fully determines  $G''(\omega)$. If only a finite set is computable, extrapolation of $|\Delta_n|^2$ to large $n$ is unavoidable.
Some intuition about the relation between  recurrents and the imaginary part of the continued fraction function at $z\to\omega+i\epsilon$  is gained by
the following examples, for complex functions $F(z)$, where $F(\omega+i\epsilon)= F'(\omega) -i F''(\omega)$.\\
(i) A Lorentzian, as given by the Drude form  (\ref{drude}),  is a {\em somewhat pathalogical limit}: All its even  moments except $\mu_0$,
are infinite.
This form  amounts to replacing the denominator's self energy by a purely imaginary constant,
\bea
F(z) &=&  \frac{1}{z-\Sigma_1(z)}\nonumber\\
\Sigma_1(z) &=&  \frac{|\Delta_1|^2 }{z-\frac{\Delta_2^2}{z-\frac{\Delta_3^2}{z- ...}}}~~\Rightarrow  ~ {i\over \tau}.
\eea
(ii)  Constant recurrents,  $\Delta_n=\Delta$, $n=1,2,\ldots$, yield a semicircle imaginary part,
\bea
F(z) &=&  \frac{1}{z-\frac{\Delta^2}{z-\frac{\Delta^2}{z- ...}}}\nonumber\\
F''(\omega) &=&  \sqrt{ 1- (\omega/2\Delta)^2  }.
\eea
(iii)  Linearly increasing
recurrents,  $|\Delta_n|^2 = n\Omega^2/2$ define the continued fractions
\beq
F(z) = \frac{1}{z-\frac{\half\Omega^2}{z-\frac{\Omega^2}{z- ...}}}
\eeq
which is relevant to HCB at high temperatures, as argued
 in Section \ref{Sec:HO}.
In Appendix~\ref{app: gaussian} it is shown that $\im F$
is a Gaussian,
\beq
 F''(\omega) =   \sqrt{\pi\over \Omega^2}  \exp\left({-{\omega^2\over
 \Omega^2}}\right).
\label{CF-gauss}
\eeq
Incidentally, the real part of $F(\omega)$ is the Dawson function~\cite{ABST}
\beq
F'(\omega) =  \frac{2}{\Omega} \int_0^{\omega/ \Omega  }
e^{t^2}dt.
\eeq
\subsection{Finite temperature corrections}
 At finite temperatures, the current fluctuations function includes the effects of the thermal density matrix
 \be
 \rho(\beta)= 2^N e^{-\beta H}/Z(\beta).
 \ee
(The prefactor of $2^N$ is introduced for later convenience).
 One can write
 \bea
 G(\beta,z) &=&   \left( \{ \rho , J_x\} ,  \cG(z) J_x  \right)_\infty\nonumber\\
 &=&   \sum_n C_n(\beta) G_n(z),\nonumber\\
 C_n  (\beta) &=& \mu_0 \left(\left\{\rho(\beta), \hat{O}_0\right\},\hat{O}_n\right)_\infty \nonumber\\
G_n(z)   &=& \mu_0\left( z- L \right)^{-1}_{n,0}.
\label{G-expand}
 \eea
Taking $z\to \omega+i\epsilon$ yields an orthogonal polynomial expansion for the current fluctuations function,
 \be
G''(\beta,\omega) =   \sum_{n=0}^\infty  C_n(\beta) P_n(\omega) G_0(\omega).
 \label{G-expand2}
\ee
The polynomials $P_n$ are orthogonal under the measure defined by
$G_0(\omega)$,
\beq
\int_\infty^\infty d\omega P_n(\omega) P_m(\omega).
G_0(\omega)=\delta_{n,m}
\eeq
In Appendix
\ref{app: orthogonal} we derive explicit expressions for $P_n$ as a function
of  $G_0$ and the preceeding  recurrents $|\Delta_m|^2,
m=1,2,\ldots  n$.

\section{Kubo formula for HCB at Half filling}
\label{sec:Kubo2}
In the previous section we introduced the moments and recurrents of the current fluctuations function $G(z)$.
Henceforth, we specialize to HCB at half filling.
First we analyze the expected asymptotic behavior of the  Liouvillian matrix elements. Second, we
construct a variational harmonic oscillator (VHO) basis in which the current fluctuations can be expanded.
Third, we generate a high temperature expansion of $G''(\beta,\omega)$.

\subsection{Liouvillian of HCB and Gaussian asymptotics}
\label{Sec:HO}
The Liouvillian hyper-operator  $\cL$  (\ref{Li}),
describes strong scattering in the following sense:  When $\cL$ acts  on a  hyperstate composed of $n$ spins,
\be
A_n  =  \sum c^{\alpha_1,\alpha_2,\ldots\alpha_n}_{i_1,i_2,\ldots i_n} S_{i_1}^{\alpha_1}S_{i_2}^{\alpha_2}\ldots S_{i_n}^{\alpha_n},\nonumber\\
\ee
the number of spins  increases or decreases by precisely one, i.e.
\be
\cL A_n  =   A_{n+1} + A_{n-1}.
\label{prol}
\ee
When $|A_{n+1}|>> |A_{n-1}|$, the primary effect of $\cL$ is to
{\em proliferate} the number of spins.

Let us compare the behavior of HCB   with  weakly interacting bosons. Consider a typical boson liquid Hamiltonian
\be
H^{weak} = a^\dagger H_0 a+ \half g a^\dagger a a^\dagger a.
\ee
The action of the respective Liouvillian on a linear operator yields
\be
[H^{weak} , a^\dagger] =  a^\dagger  H_0 + g a^\dagger a a^\dagger.
\ee
Thus, when $g$ is ``smaller'' than $H_0$, the primary effect of  $\cL$  is the first term, which {\em propagates} $a^\dagger$, rather
than the second term which {\em proliferates}  it.


The root hyperstate $\hO_0\propto J_x $, is bilinear in spins. By
repetitive applications of $\cL^n \hO_0$ one obtains clusters of
up to  $n$ spins. Consider a lattice in two dimensions or higher
~\cite{Comm:1D},  with coordination number $z > 2$. For a typical
$A_n$,  there are $n\ze$  available bonds to attach an extra spin
to an existing cluster, where  $\ze < z$. Therefore, the number of
distinct  terms  in the resulting $A_{n+1}$  is roughly a factor
of $n\ze$ more than the number of terms in $A_{n}$. We thus
expect, for such a lattice,  most of the weight of the hyperstate
$\hO_n$ to consist of  $n$ spin operators.

Since  $\hO_{n},\hO_{n-1}$ are normalized, we can crudely estimate the asymptotic  $n$ dependence  of
the recurrents
\be
|\Delta_n  |^2  \approx   \left( \hO_n, \cL \hO_{n-1}\right)^2
\sim   (4J)^2 z_{eff} n  , ~~~n>>1.
\label{asympt}
\ee
(The factor $4J$ stems from the commutation relations
$\left[S_\alpha,S_\beta\right]=i\epsilon_{\alpha\beta\gamma}S_\gamma$
and the definition of the Hamiltonian (\ref{xxz}) ). As we have
seen in (\ref{CF-gauss}), if the asymptotic relation
(\ref{asympt}) were precise, the continued fraction expansion for
$G''$ would lead to a perfect Gaussian of width $4\sqrt{2\ze}J$.

However, as demonstrated in Fig.~\ref{fig: recurrents} the
linearity of $|\Delta_n|^2$ is {\em not precise}: When $\cL$ acts
on $\hO_n$ it generates some small fraction of $n-1$ spin
operators, which {\em are not included} in the term $\Delta_n
\hO_{n-1}$ of Eq. (\ref{GS}). This spoils the exact relation between the index $n$, and the
number of  spin operators.

Nevertheless,  the coefficients $C_n(\beta)$ in
Eq.~(\ref{G-expand2}) are expected to converge rapidly at high
temperature. Expanding $C_n(\beta)$ in a high temperature series,
\beq
C_n(\beta)= \sum_i C_n^{(i)}\beta^i,
\eeq
it can be verified that the contribution of any $m$-spin  operator $A_m$ to order  $\beta^i$ depends on traces such as
\beq
\tr\left[ \left\{H^i,J_x\right\} A_m \right] =0,~~~m >  i+2.
\eeq
Therefore,
$C_n^{(i)}$ measures the relative weight of $i+2$ spins operators
or less,  in $\hO_n$. Since, as argued previously, these weights
become smaller for large $n$, we can expect for fixed $i$ that
\beq
\lim_{n >> i } C_n^{(i)} \to 0.
\label{Cni}
\eeq
{\em Thus at  a finite order  $\beta^i$, only a finite number of $C^{(i)}_n$'s are of substantial magnitude.}

Phrased differently: In an idealized situation in which $\hO_n$
would contain  only products of  $(n+2)$-spins, the $C_n$'s would
decay with $\beta$ as
\be
C_n = {1\over Z} \tr \left(  e^{-\beta H} \{ \hO_n, \hO_0\}
\right) \sim \beta^{n}C_n^{(n)}+o(\beta^{n+2}).
\label{HiT-ideal}
\ee

This discussion  raises  an obvious question: {\em Is there a lattice  where Eq.~(\ref{HiT-ideal}) becomes precise?}.  We have preliminary expectations \cite{wDan}, that
this would be the case, at least for low orders in $n$, for the Bethe lattice in the limit of large coordination number. The rapid convergence of the harmonic oscillator basis discussed below,
indicates that the square lattice at high temperatures is not ``far'' from the infinite dimensional limit.

\subsection{Variational harmonic oscillator  expansion}
\label{sec:VHO}
A high temperature expansion for the
conductivity can be generated by choosing a convenient basis to expand
$G''(\omega)$. The linear increase of the recurrents, suggested
earlier in Eq. (\ref{asympt}) implies a Gaussian decay of
$G''(\omega)$ at high frequencies. Therefore it is natural to
choose a variational harmonic oscillator (VHO)  basis such that,
\be
\tilde{G}''(\beta,\omega) =  \sum_{n=0}  D_n(\beta)  \tilde{H}_n(\omega) \psi^2_0(\omega),
\label{expand}
\ee
where $\tilde{H}_n(\omega)=H_n(\omega/\Omega_v)/\sqrt{2^n
n!}$, and
\beq
\psi^2_0 (\omega)  =      {1 \over  \sqrt{\pi \Omega^2_v}}  \exp\left( -  {\omega^2 \over  \Omega_v^2 } \right).
 \label{g.s.}
\eeq
The function $\psi^2_0(\omega)$ is the VHO ground state
probability density, with a variational frequency scale
$\Omega_v$, and $\tilde{H}_n(\omega)$ are the corresponding
(normalized) Hermite polynomials, which constitute an orthogonal
set under the measure of $\psi_0^2(\omega)$.

The moments of $\tilde{H}_n(\omega)\psi_0(\omega)^2$ are
\bea
\lambda_n^k &=&    \int_{-\infty}^\infty {d\omega\over \pi} \omega^k  H_n(\omega/\Omega_v) \psi^2_0(\omega)\nonumber\\
&=&  \sqrt{\frac{2^n}{n!}}2^{-k}~(\Omega_v)^{2k}
\frac{(k/2)!}{((k-n)/2)!}\Gamma\left(k+{1\over 2}\right),
\label{lambda}
\eea
for $k \ge n$. By construction, all $\lambda_n^k$ vanish for
$n>k$.

Using the high temperature expansion of a finite number of moments
\be
\mu_k(\beta)= \sum_{i=0}^\infty \mu_k^{(i)} \beta^i,     ~~~~~k=0,2,\ldots n_{max},
\label{muk-hiT}
\ee
and similarly, 
\be
D_n(\beta)= \sum_i D_n^{(i)}\beta^i,~~~~n\le n_{max}.
\label{Cn-hiT}
\ee

We insert (\ref{expand}) into
(\ref{moments}) and obtain  a finite set of
linear equations for each  $D^{(i)}_n$,
\be
\sum_{n=0}^{n_{max}} \lambda_{n}^k D^{(i)}_n  = \mu^{(i)}_k(\beta),~~~k=  0,2,\ldots n_{max}.
\label{invert}
\ee
Solving for (\ref{invert})  yields the desired coefficients  for Eq. (\ref{expand}).

\begin{figure}
\vspace{-0.3cm}
\begin{center}
\epsfxsize=.60\textwidth \centerline{\epsffile{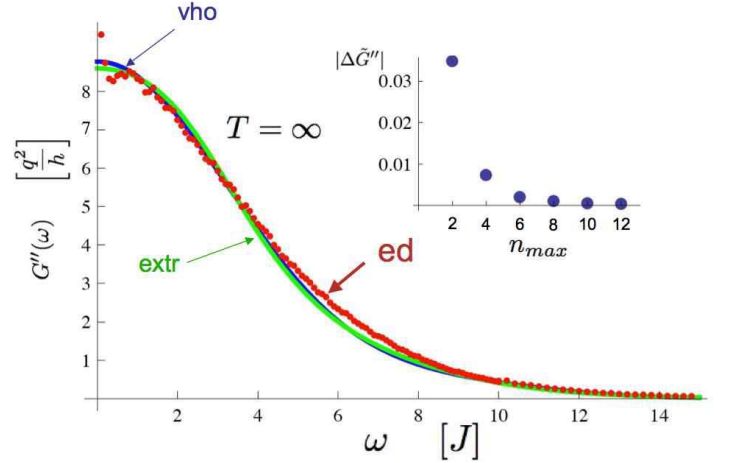}}
\vspace{-0.5cm}
\caption{ Infinite  temperature current fluctuations function  ${G_\infty ''}(\omega)$.
Solid lines (blue online) {\bf vho} depicts the results of the  variational
harmonic oscillator expansion, and (Green online) {\bf extr} depicts
the result of the reccurrents extrpolation, shown in Fig.
\ref{fig: recurrents}. Circles (red online) {\bf ed} depict  the
exact diagonalization result computed on a $4\times 4$ lattice. Inset: Convergence
of the  variational harmonic oscillator expansion. $|\Delta \tilde{G}''|$ are the distances
between consecutive approximants, and $n_{max}$ is the number of computed moments.}
\label{fig:G-om}
\end{center}
\end{figure}
\section{Dynamical conductivity:  high temperature}
\label{sec:DCHT}
The dynamical conductivity, to leading order in $\beta$, is
\be
\sigma_{\beta\to 0} = {\beta \over 2} G''_{\infty}(\omega).
\ee
We have  calculated the infinite temperature current fluctuation $\tilde{G}''_{T=\infty}$, using three distinct methods.
The results of the three approaches show satisfactory agreement,  as depicted in Fig.\ref{fig:G-om}.
First we computed the infinite temperature moments, listed in Table \ref{table:muk0},
\be
\mu^\infty_k={2\over 2^N}\tr  \left(J_x  \cL^k J_x\right), ~~~k=0,2,\ldots
12.
\label{muk}
\ee
These traces were computed numerically on a
finite 16 site square lattice with periodic boundary conditions.
At  $k\ge 8$ the calculation introduces finite size errors,
due to loops of operators which circulate around the system.  We
eliminated most of the  contributions of these loops  by averaging over
Aharonov-Bohm fluxes through two holes of the torus~\cite{LAA}.

\begin{table}
\begin{tabular}{|c|c|}
\hline
$k$ & $\mu_k^{\infty}$ \\
\hline
0 & 4 \\
2 & 64 \\
4 & 4096 \\
6 & 544768 \\
8 & $1.20906 \times 10^8$ \\
10 & $3.96113 \times 10^{10}$ \\
12 & $1.75571 \times 10^{13}$ \\
\hline
\end{tabular}
\caption{Low order moments at $T=\infty$.}
\label{table:muk0}
\end{table}

\begin{figure}
\vspace{-0.3cm}
\begin{center}
\epsfxsize=.55\textwidth \centerline{\epsffile{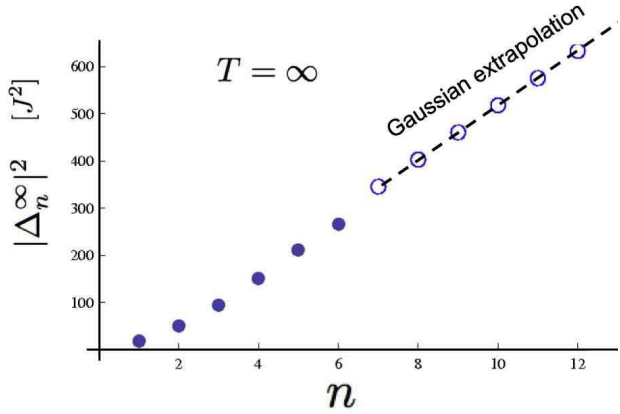}}
\vspace{-0.5cm}
\caption{The recurrents of $G_\infty''(\omega)$ at infinite temperature, and their extrapolation to large index $n$.}
\label{fig: recurrents}
\end{center}
\end{figure}


The VHO calculation of $G''$  follows Section \ref{sec:VHO}. Inverting Eq.
(\ref{invert}), for the $n_{max}$ lowest moments,  we obtain
$\tilde{G}_{n_{max}}''$. We obtain the  variational frequency
\be
\Omega_v=7.48 J,
\ee
which  minimizes the distance  between the two highest order approximants,
\beq
|\Delta \tilde{G}''|^2(\Omega_v)  =
\frac{\int_{-\infty}^\infty d\omega \left| {\tilde{G}_{n_{max}}''}
- {\tilde{G}_{n_{max}-2}''}\right|^2}{\int_{-\infty}^\infty
d\omega \left| \tilde{G}_{n_{max}-2}''\right|^2}.
\eeq
We plot $\tilde{G}''$ for  $n_{max}=12$  as the solid (blue
online) curve in  Fig.~\ref{fig:G-om}.
 In the inset of Fig.~\ref{fig:G-om}, we plot the convergence as a function of $n_{max}$. The rapid decay of $\Delta \tilde{G}''$ indicates convergence
 of the VHO expansion at $T=\infty$.

The second calculation extrapolates the recurrents, as shown in Fig.~\ref{fig: recurrents}.
We compute  $|\Delta^\infty_n|^2,~~n=0,1,\ldots 6$, from the known moments,  using   (\ref{mom-rec2}).
The higher order  recurrents exhibit an approximate linear
increase with $n-1$ which is consistent with a Gaussian decay of
$G''$ at high frequencies as given  by the continued fraction example of Eq.~(\ref{CF-gauss}). We
continue the linear slope (see dashed line in Fig.~\ref{fig:
recurrents}) by the extrapolation,
 \be
|\Delta^\infty_{n>6}|^2   \to   z_{eff} (4J)^2 (n-1), ~~~~~
z_{eff}=3.59,
\label{lin-rec}
\ee
 where $z_{eff}$  is the   ``effective coordination number''.
The approximate function ${\tilde{G}''}_{extr}$  is obtained using
the continued fraction representation (\ref{CFinfty}), and plotted
as the solid (red online) curve in Fig.~\ref{fig:G-om}. We notice
that the two approaches yield very similar curves.

Last, we compute  the  infinite temperature current fluctuations function by exact diagonalization (ED)
in the Lehmann representation,
\beq
\tilde{G}''_{ed}(\omega)=  \frac{2\pi }{2^N}\sum_{n,m}|\6n|J_x|m\9|^2\delta(\omega+E_n-E_m).
\label{G0L}
\eeq
where $|n\rangle, E_n$ are the  eigenstates and eigenspectrum of
$H$ on a $4\times 4$ lattice with periodic boundary conditions. We
expect the finite size effects to be small at infinite
temperatures, where the correlation length is much shorter than
the lattice size.  Indeed, the agreement between the three
approaches shown in Fig.~\ref{fig:G-om},  supports  this
expectation.

\begin{figure}
\vspace{-0.3cm}
\begin{center}
\epsfxsize=.60\textwidth \centerline{\epsffile{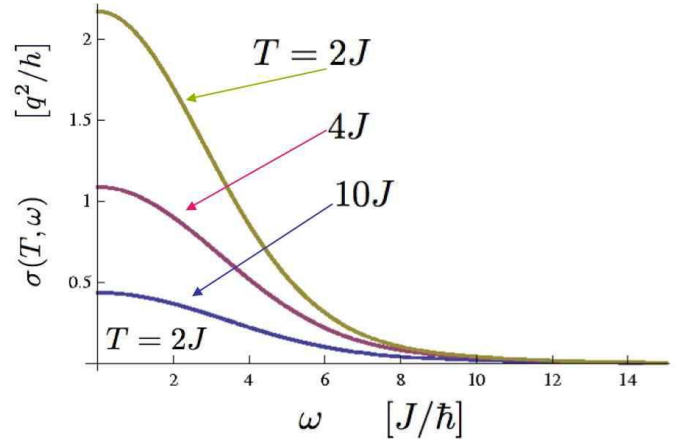}}
\vspace{-0.5cm}
\caption{Dynamical  conductivity in the high temperature resistive phase. The function can be fit by Eq. (\ref{sigma-app1})}
\label{fig:sigma}
\end{center}
\end{figure}


\begin{figure}
\vspace{-0.3cm}
\begin{center}
\epsfxsize=.55\textwidth \centerline{\epsffile{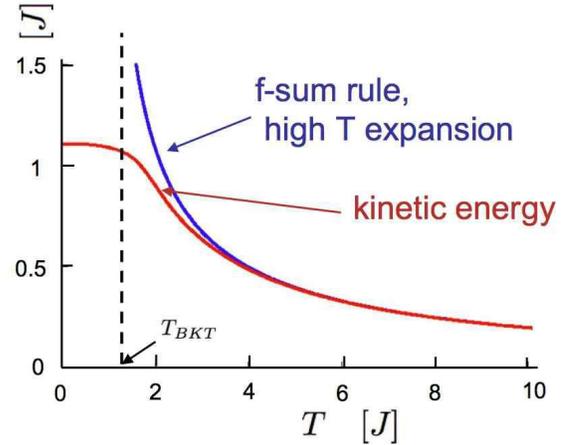}}
\vspace{-0.5cm}
\caption{The temperature dependent kinetic energy $\6 - K_x \9$ (solid, red online), calculated by Refs.~\cite{ding, sandvik,rogiers} and the sum rule
for the high temperature expansion of the dynamical conductivity of Fig.~\ref{fig:sigma},
(solid, blue online). The vertical dashed line  marks the superconducting phase below $T_{BKT}$. }
\label{fig:sumrule}
\end{center}
\end{figure}


\subsection{Finite temperature corrections and f-sum rule}
Order  $\beta^2$  corrections to $G''$ are obtained by a high
temperature expansion of the moments  $\mu_k^{(2)}$ as defined in
Eq.~(\ref{muk-hiT}). Inverting Eq.~(\ref{invert}),  the
coefficients $D_n^{(2)}$ were  computed up to order $n_{max}=12$,
and inserted into Eq. (\ref{Cn-hiT}). The resulting  temperature
dependent dynamical conductivity is plotted in
Fig.~\ref{fig:sigma}, for several temperatures. A crude analytical
approximation is given by
\bea
\sigma(T,\omega) & \approx & 0.91 {q^2 \over h   }
{\tanh(\omega/(2T))\over (\omega/\bar{\Omega})} e^{-(\omega/\bar{\Omega})^2},\nonumber\\
\bar{\Omega} &=&  4.8 J.
\label{sigma-app1}
\eea

The truncation  errors in the VHO expansion,  and the high temperature series, are monitored
by comparing the integrated conductivity to the kinetic energy, using the f-sum rule equation
\beq
\int_{-\infty}^{\infty}{d\omega\over \pi} {\tanh(\beta\omega/2)\over \omega} G'' (\beta,\omega) =  \6 -q^2 K_x \9_\beta.
\label{fsumrule-HT}
\eeq
The values of  $\6  K_x \9_\beta$  were  computed by high
temperature series expansion up to 11th order  in  Ref.~\cite{rogiers}, and at low temperatures by quantum Monte-Carlo simulations~\cite{ding,sandvik}.
In Fig~\ref{fig:sumrule} we plot the f-sum rule,
Eq.~(\ref{fsumrule-HT}), with $G''$ calculated to order $\beta^2$,
and the temperature dependent kinetic energy. The  corrections
grow larger than 10\% at  $T < 2T_{BKT}$.

We note, however, that numerical satisfaction of Eq.~(\ref{fsumrule-HT}) does not, in general, ensure the
accuracy of the DC conductivity $\sigma(0)$.  In fact, there is an expectation of a
cusp at zero frequency arising from high order 
non-linear coupling between the current and other long lived 
diffusive modes \cite{MOH}. The magnitude of this feature is expected to be weak 
because of particle hole symmetry.  From our calculations we can conclude that it
is below the  resolution provided by the exact diagonalization and the 12th order moment expansion \cite{HusePC}.

  \section{DC Resistivity: High temperature}
  \label{sec:DC}
Based on the calculations of $\sigma(T,\omega)$ of the previous section,  the high temperature expansion of
the DC resistivity, shown in Fig.~\ref{fig: resistivity},  is given to order $1/T$ as
\beq
R(T) = 0.23~R_Q { T \over  J } \left( 1-2.9(J/T)^2
+\cO((J/T)^4),\right)
\label{rxxT}
\eeq
where $R_Q=h/q^2$ is the boson quantum of resistance.
In the regime of $T\ge 2T_{BKT}$, the negative corrections of order $T^{-3}$
are relatively small, as the sum rule shows in Fig.~\ref{fig:sumrule}.

As   $T_{BKT}$ is approached from above the resistivity drops rapidly.
The critical regime  was described by
Halperin and Nelson
~\cite{HN,AHNSprl,AHNS} (HN) who considered the contribution of unbound vortices
to the  charge transport coefficients.  Using  vortex-charge duality, and Einstein's
relation for vortex conductivity, HN derived the critical resistivity as
\beq
R^{HN}= R_Q  {h n_v D_v \over T},
\label{HN}
\eeq
where $D_v$ is  vortex  diffusion constant, and $n_v$ is the density of free vortices.
Estimating that $n_v = \xi(T)^{-2}$ using  (\ref{xiT}),   the critical resistivity is expected to be suppressed toward $T_{BKT}$ as
\beq
R^{HN}(T) = R_Q { h D_v n_0\over T} A^{-2} \exp\left( -2 B\left({T_{BKT}\over T-T_{BKT} }\right)^{1/2}\right),
\label{eq:nf}
\eeq
where $n_0$ is a microscopic density of order $a^{-2}$.

The precise diffusion constant for the HCB model is not known, and can be expressed as
\be
D_v n_0 =   \kappa  J/\hbar,
\label{kappa}
\ee
where $\kappa$ is an undetermined dimensionless constant. We can
estimate the numerical value of $\kappa$ by requiring that
$R^{HN}(T)$ matches  Eq.~(\ref{rxxT}) at  $T \ge  2 T_{BKT}$ where
the f-sum rule is satisfied to a higher accuracy. We find that
matching occurs in the range,
\be
0.38 < \kappa < 0.51.\label{kapp}
\ee
We can interpret the  vortex diffusion constant $D_v= l^2 /\tau$,
as arising from a scattering time of order $\tau=\hbar/J$ (the
inter-site vortex hopping time, according to calculations of Ref.
~\cite{LAA})  and a short mean free path $l\approx a$.  Eqs. (\ref{kappa}),(\ref{kapp}) are consistent with the value $D_v\approx \hbar/m$
which was posited~\cite{AHNSprl} for helium films.

\section{Dynamical Conductivity: zero temperature}
 \label{sec:T=0}
At $T=0$ the conductivity is given by
\be
  \sigma_0 (\omega)=q^2 \pi \rho_s\delta(\omega) + {G''_{T=0} (\omega) \over  |\omega|}.
  \ee

   We calculate the current fluctuations $\tilde{G}''_{T=0}$ in two stages. First, we appeal to the relativistic Gross Pitaevskii field theory~\cite{fisherlee}, and obtain the low frequency gap and threshold form of the function.
 The mass parameter $m$, and the high frequency Gausssian scale $\Omega_0$
 are variational fitting parameters.
Second, we compute  the lowest $12$ moments by exact
diagonalization of a  20 site cluster. These determine the lowest
recurrent parameters $|\Delta_n^0|^2$ which are fit to the
variational form.
\begin{figure}
\vspace{-0.3cm}
\begin{center}
\epsfxsize=.50\textwidth \centerline{\epsffile{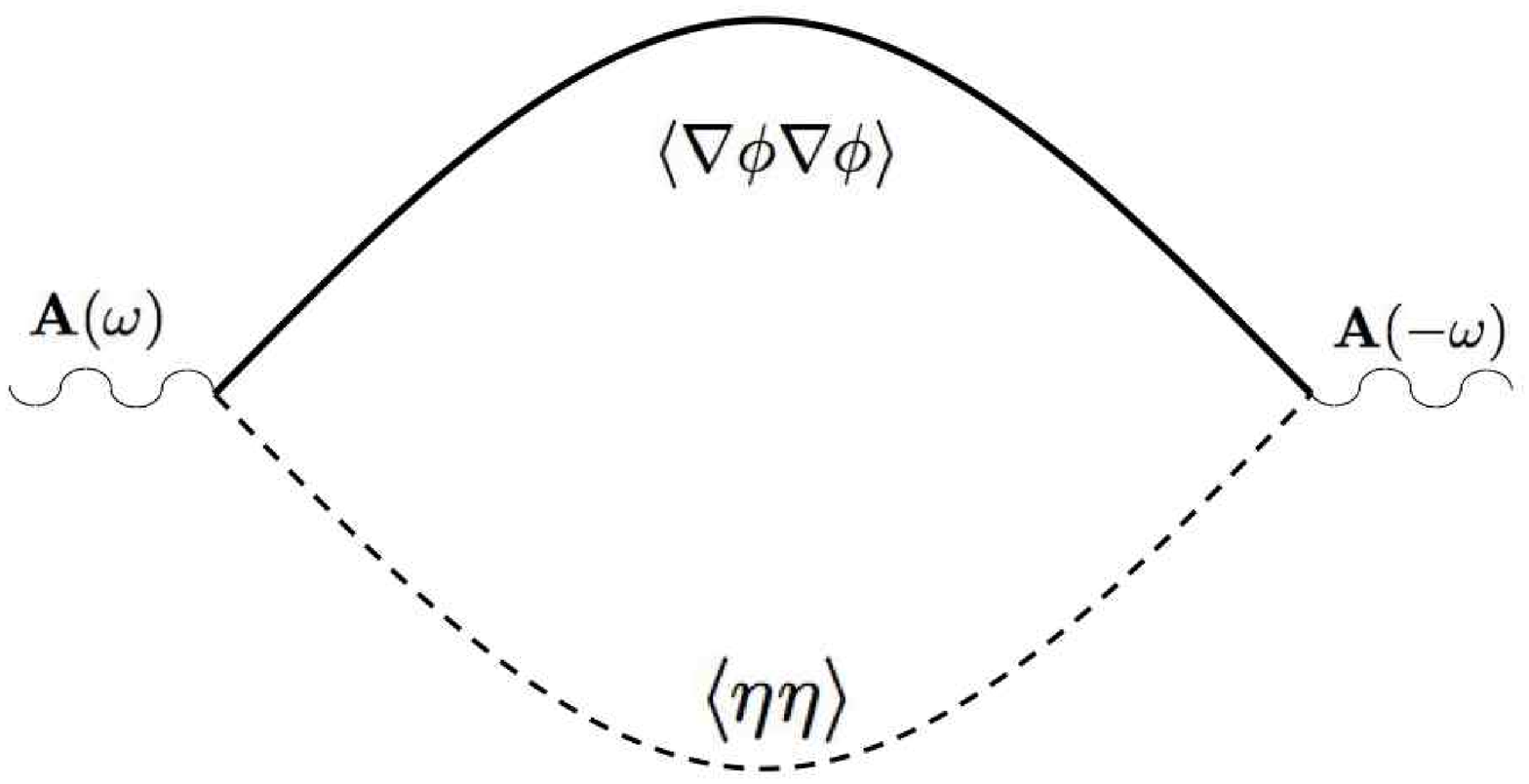}}
\vspace{-0.5cm}
\caption{Low order current fluctuations function in the relativistic Gross Pitaevskii field theory,
as calculated by Eq. (\ref{bubble-diag}).
Dashed line describes gapped magnitude fluctuations. Solid line describes gapless phase fluctuations. }
\label{fig:bubble}
\end{center}
\end{figure}


\subsection{Field theoretical calculation}
The  relativistic Gross-Pitaevskii (RGP) model describes the long
wavelength theory of quantum rotators in the absence of a linear
time derivative term~\cite{fisherlee}. We use it to describe the HCB action
at half filling, which exhibits particle-hole symmetry.
$\Psi(\br,\tau)$ denotes the fluctuating condensate order
parameter, governed by the imaginary time action,
\beq
S_{RGP}=\int d^2x\int d\tau  \half |\dot{\Psi}|^2 +  {\rho_s\over 2\Delta^2}|\nabla \Psi|^2 -{m\over 8\Delta^2}\left( |\Psi|^2 - \Delta^2\right)^2,
\label{RGP}
\eeq
where $\Delta$ is the ground state order parameter, and $m$ is the effective (Higgs) mass.
Expanding  (\ref{RGP}) to second order about $\Psi=\Delta(1+ \eta)e^{i\phi} $, the harmonic fluctuations
are described by decoupled phase and magnitude modes,
\bea
S_{RGP}^{(2)}&=&\half  \int d^2x\int d\tau   \left(\dot{\phi}^2 \Delta^2 +  \rho_s |\nabla \phi|^2\right) \nonumber\\
&&+\half   \int d^2x\int d\tau \Delta^2 \left( \dot{\eta}^2  +{  \rho_s\over \Delta^2} |\nabla \eta|^2 +  m \eta^2 \right) .\nonumber\\
\label{S2}
\eea
 We introduce a vector potential by shifting $\nabla \to
\nabla+iq\bA$. The current operator is obtained from $J_x=\delta S/\delta
A_x$, therefore its paramagnetic part, expanded to the same order
as Eq.~(\ref{S2}), is given by
\beq
J_x(\bx)=q\rho_s \left(\nabla\phi + 2\eta\nabla\phi\right).
\eeq
The lowest order current fluctuations function is given by the
bubble diagram~\cite{subir,cha} depicted in Fig.~\ref{fig:bubble},
\bea
&&G_{RGP}(i\omega_n)\simeq
\frac{4q^2}{\beta}\int \sum_{\nu_n}  {d^2\bk\over (2\pi)^2}     \nonumber\\
&&~\times {k_x^2 \over \left(c_s^{-2}(\omega_n+\nu_n)^2+(\bq+\bk)^2+c_s^{-2}m^2\right) \left(c_s^{-2}\nu_n^2+\bk^2\right) },\nonumber\\
\label{bubble-diag}\eea
where $c_s=\sqrt{\rho_s/\Delta^2}$ is the speed of sound. We compute
the integral by performing the Matsubara sum, and take $\bq \to 0$
and $\beta\to
\infty$, which yields
\be
G''_{RGP}(\omega) \simeq
\frac{q^2}{4}\left(\frac{\omega^2-m^2}{2\omega}\right)^2\frac{1}{|\omega|}\Theta(|\omega|-m).
\label{eq:
diagram result}
\ee
The dynamical conductivity therefore exhibits finite frequency weight above the mass gap
 $m$. We cannot rule out that higher  order
interactions may produce subgap spectral weight, since magnitude
excitations can decay into  phason pairs ~\cite{OSF-Theory}.

In addition, the RGP theory   Eq.~(\ref{RGP}) only describes
long wave length  fluctuations. Therefore,   lattice scale and high energy  cut-off effects are important
for the high frequency tails of $G''$.  These will be computed
directly from the zero temperature Kubo formula.

\begin{figure}
\vspace{-0.3cm}
\begin{center}
\epsfxsize=.55\textwidth \centerline{\epsffile{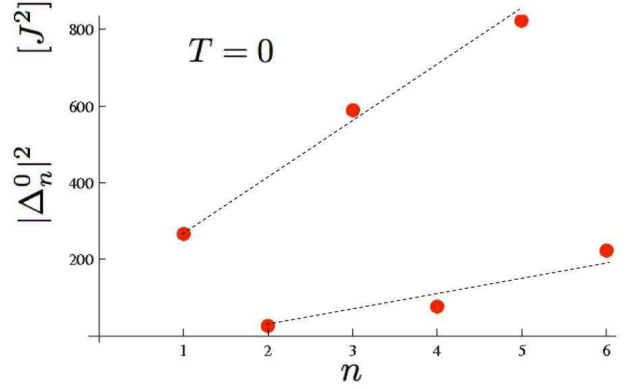}}
\vspace{-0.5cm}
\caption{The recurrents of $G_{T=0}''(\omega)$ at zero temperature. Dashed line
emphasize the different behavior of even and odd recurrents, which is related to the gap structure in Fig.~\ref{fig: ground state}.  }
\label{fig:recurrents_gs}
\end{center}
\end{figure}


\subsection{Variational Fit of Recurrents}

The moments of $G''_{T=0}$, Eq.~(\ref{moments}), are
equal to the ground state expectation values,
\be
 \mu_{k} = \langle 0|  \left\{ J_x, \cL^k J_x \right\}|0 \rangle.
\label{GSmoments}
\ee
 We compute  a set of  $\mu_{k}, k \leq 12$ by exact diagonalizations on a $5\times 4$ lattice. Using Eq.~(\ref{mom-rec2}) we determine the set of recurrents
 $|\Delta_n|^2, n=1,2,\ldots 6$, which are depicted in Fig.~\ref{fig:recurrents_gs}.
Finite size effects are not expected to be large for small $n$,
where $\Delta_n$ depend mostly on short range correlations. We notice a
striking difference between even and odd recurrents, which seem to
follow {\em two} linearly increasing slopes. This even-odd effect
of the recurrents, is an indicator (not a proof) of a gap-like
structure~\cite{muller}.

Motivated by the field theoretical calculation, we  use the trial function,
\be
\tilde{G}''_{T=0} \propto G''_{RGP}(\omega)  \exp(-\omega^2/\Omega_0^2),
\ee
where  $m$ -  the mass gap, and  $\Omega_0$ - the high freqeuncy
fall off, are variational parameters. We use (\ref{mom-rec2}) to
determine the trial recurrents. The variational parameters are
then determined by a least squares  fit  between  the exact and trial  recurrents.

\begin{figure}
\vspace{-0.3cm}
\begin{center}
\epsfxsize=.62\textwidth \centerline{\epsffile{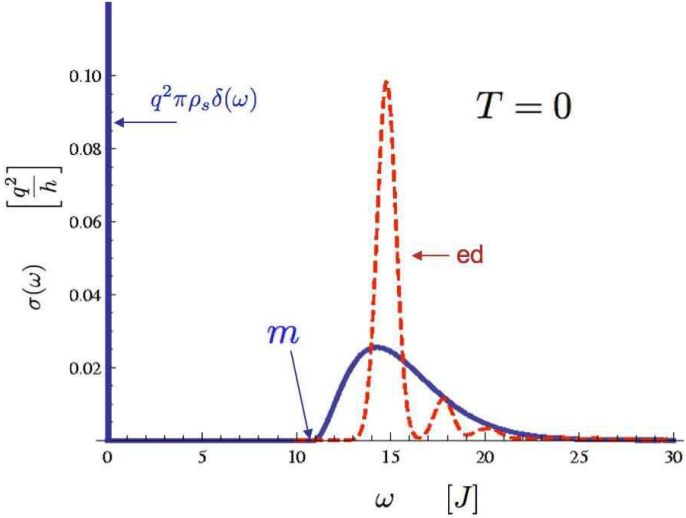}}
\vspace{-0.5cm}
\caption{Zero temperature dynamical conductivity of HCB.
The Higgs mass gap $m$ is depicted.
 Solid line (blue online)
is determined by variationally  fitting the recurrents to those in Fig.~\ref{fig:recurrents_gs}.   Dashed line (red online)
is computed by exact diagonalization of a 20 site cluster, with $\delta$-function broadening of $0.5J$. }
\label{fig: ground state}
\end{center}
\end{figure}


Thus we obtain,
\bea
\Omega_0 &\to & 8.6 J,  \nonumber\\
m &\to & 10.9 J.
\eea
The mass gap appears to be similar to the  high frequency Gaussian fall-off of the conductivity
at high temperatures, as given by the linear slope of the
recurrents, Eq.~(\ref{lin-rec}).

The resulting dynamical conductivity  is  depicted in
Fig~\ref{fig: ground state}. The figure also includes, for
comparison, the results of  exact diagonalization on a 16 site
cluster, as given, for $\omega>0$ by
\beq
\tilde{G}''_{ed}(\omega) = \pi  \sum_{m}|\60|J_x|m\9|^2\delta(\omega+E_0-E_m).
\label{G0L}
\eeq
The oscillations in $\tilde{G}''_{ed}$ are artifacts of finite size gaps in the spectrum.
We see that the two curves agree on the position of the central peak
and the total spectral weight.

A test of these calculations is provided by
the zero temperature f-sum rule is,
\be
{h\over q^2}\int_{0+} { d\omega\over \pi}  {G_{T=0}''(\omega) \over \omega}= \langle -K_x\rangle -\rho_s,
\ee
The left hand side for the varational and  exact diagonaliozation results yields
\be
{h\over q^2}\int_{0+} { d\omega\over \pi}  {\tilde{G}_{T=0}''(\omega) \over \omega} =  \left\{ \begin{array}{ll}
0.0148~J & \mbox{Variational}\\
0.0164~J & ED
 \end{array} \right..
\ee
The sum rules are comparable in magnitude to  values obtained by QMC ~\cite{sandvik} for the zero temperature kinetic energy and superfluid stiffness,
\bea
\langle -K_x\rangle &=& 1.09765(4)~J,\nonumber\\
\rho_s &=&   1.078(1)~J,\nonumber\\
\leadsto~\langle -K_x\rangle -\rho_s &=& 0.019(1)~J.
\eea

We notice the very small spectral weight (2\%)  of the high frequency peak at zero temperature, relative
to the condensate weight. This weight is due to the quantum fluctuations of the ground state, and to the
non  conservation  of the current operator in two dimensions. This weight can be ascribed to the magnitude oscillations mode.

{\em Note:} An important
question is whether the mass gap survives corrections to Eq. (\ref{bubble-diag}).
Within our variational
approach, we tried to answer this question by allowing sub-gap
spectral weight parameterized by a power-law tail such as
\be
\tilde{G}''_{T=0} \propto \left( {  (|\omega| / m)^\alpha \over 1+ (|\omega|/m)^\alpha}\right) \exp(-\omega^2/\Omega_0^2).
\ee
The least squares  fit of the recurrents has found that  $\alpha$ tends to increase indefinitely.
This is consistent, (although not being a  proof)   with having a true gap at $\omega=m$.


\section{Discussion}
\label{sec: discussion}
\subsection{Bad Metallicity}
HCB  at half filling exhibits {\em ''bad
metal''} characteristics, as  demonstrated  in Eq. (\ref{sigma-app})
and in Fig.~\ref{fig: resistivity}.

In contrast to  conventional metals and bosonic gases at high temperatures ~\cite{res-sat1,res-sat11,res-sat2},
the resistivity of HCB  rises approximately linearly, without a sign of
saturation at $R\approx h/q^2$.
Such behavior  signals the breakdown
of   Boltzmann equation, since the mean free path becomes shorter than interparticle distance~\cite{badmetal}.

A related quantity is the width of the low frequency conductivity
peak, which in metals is called the ''Drude peak''.
 Here, Eq.
(\ref{sigma-app}) shows that the low freqeuncy temperature
dependent peak in $\sigma(\omega)$ is goverened mostly by the
fluctuations-dissipation factor~\cite{jaklic}.  If   one   fits
the width by Eq. (\ref{drude}) in the regime $T_{BKT}< T <<
\bar{\Omega}$, one obtains,
\be
\left( {1\over \tau}\right)^{eff} = 2T.
\ee
We note however, that for our hamiltonian (\ref{xxz}),   $T_{BKT}$
is not much smaller  than $\bar{\Omega}$. Therefore, separating
the two frequency scales in $\sigma(T,\omega)$ is difficult.
However,  the ratio of the two scales can made larger by
additional interactions.

The HCB model relates  the asymptotic  resistivity slope to the
zero temperature superfluid stiffness:
\be
 {dR^\infty \over dT} =  0.245  {R_Q \over \rho_s}.
\label{hl0}
\ee

In a three  dimensional  system  of weakly coupled layers,  the
transition temperature $T_c$ is shifted from $T_{BKT}$ by a factor
given by Eq. (\ref{T3D})~\cite{HT,Mihlin}.  Above $T_c$ the density of free vortices rises rapidly, which causes the rise in $R(T)$,
as shown in Fig.~\ref{fig: resistivity}. We can operationally define a nominal {\em critical
normal state resistivity} $R_c$ by
\be
R_c \equiv   {dR^\infty \over dT} T_c.
\ee
Thus, a HCB version of ''Homes law'' is obtained:
\be
\rho_s(0)  =   0.245   {R_Q\over R_c}    T_c.
\label{homes}
\ee

\begin{figure}
\vspace{-0.3cm}
\begin{center}
\epsfxsize=.64\textwidth \centerline{\epsffile{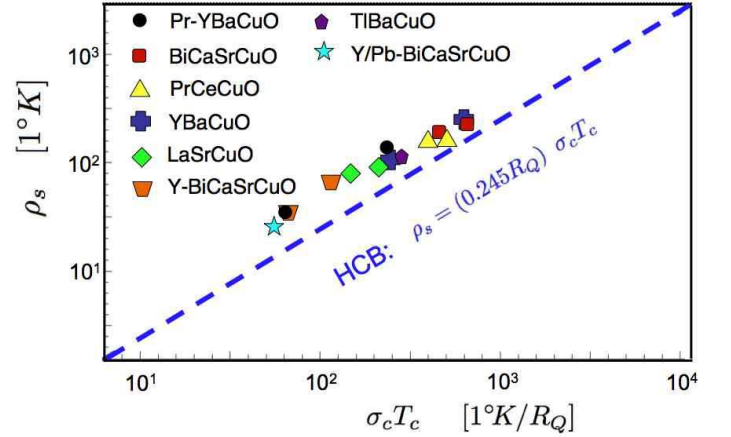}}
\vspace{-0.5cm}
\caption{Comparison of ''Homes law'' of  hard core bosons (HCB) and cuprate supeerconductors.  HCB calculation is depicted by
the dashed line (blue online). 
Cuprate data, compiled from Refs.~\cite{homes,homes-PRB},  describes
the two dimensional critical conductivity $\sigma_c$, as measured by optical conductivity exprapolated to zero frequency, at the transition temperature $T_c$.
$\rho_s$ values are compiled from measured plasma frequencies by Eq.~(\ref{rhos-ops}).
$R_Q=6453 \Omega$ is the boson quantum of
resistance. Different data points
of the same symbol  correspond to different doping concentrations of the same compound.
}
\label{fig:homes}
\end{center}
\end{figure}


\subsection{Cuprate Conductivity}
 A large linear in $T$
resistivity~\cite{linear} and
optical relaxation rate~\cite{Basov-Timusk, Basov-Timusk2} are
widely observed in  clean  samples, especially  near optimal doping. It has been shown~\cite{Chamon}
that the  linear resistivity is not consistent with a proximity to  a quantum critical point.

It is plausible therefore, that the ''bad
metal'' characteristics of the normal phase of cuprates may be
described  by lattice bosons in their resistive state. Support to this viewpoint is given by 
Uemura's empirical scaling law $T_c \propto \rho_s^{ab}(T=0)$
~\cite{Uemura}, and the observation of a superfluid density jump in
ultrathin underdoped cuprate films~\cite{XYjump}. These are consistent
with the behavior of a bosonic superfluid, captured by an
effective $XY$ model. In underdoped cuprates, additional evidence
exists that the hole pair bosons  survive above $T_c$, up to the
pseudogap temperature scale~\cite{pairing,yazdani} $T^*>T_c$. Thus there
have been several theoretical approaches~\cite{Ranninger,  LB-cuprates,PBFM, LB2,Mihlin}
to the superconducting properties of cuprates based on lattice
bosons of charge $q=2e$.

We note
however, that above the pseudogap temperature and frequency scale $T^*$,
hole pairs completely disintegrate. Thus, {\em above $T^*$,  the HCB model cannot
be used to obtain the  temperature and frequency dependence
correctly.   }

Homes {\em et al.}~\cite{homes}, have pointed out, that the superfluid
stiffness  is generally proportional, with a seemingly universal constant,
to the product of  ''critical  conductivity'' $\sigma^{3D}_c$,
times $T_c$. The ''critical  conductivity'' $\sigma_c$   was experimentally  defined by extrapolating $\sigma(\omega,T_c)$  to zero frequency.
This empirically universal scaling  was interpreted as a sign of ''dirty superconductors'', where
$T_c$ is determined by the disorder driven  scattering rate~\cite{homes-PRB}.

Here we promote a different viewpoint, which maintains that  Homes law can also be obtained in a {\em disorder free } model.
The necessary ingredients are strong scattering effects, as described by hard core  bosons above $T_{BKT}$.  

To relate the experimental data of cuprates to
the HCB model, we first translate the 3D critical conductivity to
a two dimensional critical 
conductivity:
\be
\sigma_c = a_c \sigma_c^{3D},
\label{sig3D}
\ee
where $a_c \approx 1.5\mbox{nm}$ is the  interplane distance.

The
zero temperature boson superfluid stiffness,  can be deduced from
the in-plane  London penetration
depth,
\be
\rho_s= a_c  \rho_s^{3D} = \frac{\hbar^2c^2}{16\pi
e^2}\frac{a_c}{\lambda_{ab}^2},
\ee
where $c$ is the speed of light, or the optical  plasma frequency
$\omega_{ps}$:
\be
\rho_s =  \frac{\hbar^2}{16\pi
e^2}a_c\omega_p^2.
\label{rhos-ops}
\ee
In Fig.~\ref{fig:homes} we plot the data reported in  Ref.~\cite{homes,homes-PRB}
by translating it into the two dimensional quantities. We find, quite remarkably, that  the
slope of Eq. (\ref{homes}) lies quite close to the
experimental data. We emphasize that the proportionality constant in Eq. (\ref{homes}) is not expected to
be universal. A priori, one might expect it to vary with additional interactions and the filling.
If the agreement with cuprates' data is not fortuitous,  it would imply that this  constant might not be sensitive to moderate variations of $H$.

At very low temperatures, we find that there is a signature of the
quantum fluctuations of the order parameter in the optical
conductivity at high frequencies. These are magnitude fluctuations
characterized by a ''Higgs mass'' gap at frequency
\be
m\approx 10 \rho_s.
\ee
It is interesting to mention that such a massive magnitude mode,
given by the relativistic Gross-Pitaevskii
action, appears in the superfluid phase of the strongly interacting Bose Hubbard model
at integer filling. It has been associated with the  ''oscillating
superfluidity'' experiments of cold atoms in an optical lattice
~\cite{OSF1,OSF2,OSF-Theory}.

 While a clear signature of such  a peak in cuprates has not been identified, we speculate that
it might be related to  the ubiquitous  {\em mid infrared peak}
which has been detected in several compounds  at low temperatures
~\cite{MIR}. We caution however, that at these high
frequencies, additional fermionic excitations  become increasingly  important.

In summary, we conclude that some of the ''normal state''
phenomenology of  cuprates, and perhaps other unconventional
superconductors, may be described by lattice bosons. However, we
emphasize that  Eq. (\ref{xxz}) oversimplifies these systems, by
omitting potentially important ingredients: fermionic excitations,
long range Coulomb interactions, interlayer coupling, disorder
and inhomogeneities, and of course the HCB conductivity
away from half filling. Clearly, these effects need to be
accounted for in future work.

\section{Acknowledgments}
We thank Dan Arovas,  Yosi Avron,   Dimitri Basov, George
Batrouni,  Christopher Homes,  David Huse, Peter Jung,  Steve Kivelson, Daniel Podolsky and
Efrat Shimshoni, for useful discussions. We acknowledges support
from Israel Science Foundation, and US - Israel Binational Science
Foundation. AA is grateful for Aspen Center for
Physics which inspired some of the ideas in this paper. NL
acknowledges the financial support of the Israel Clore foundation.

\appendix
\section{Orthogonal polynomials and recurrents}
\label{app: orthogonal}
Iterative use of the matrix inversion formula for $G_n(z)$
\beq
G_n(z)=  \mu_0\left( z- L \right)^{-1}_{n,0},
\eeq
with $L$ defined as in (\ref{eq: cLmat}) yields, for
$z\to\omega+i\epsilon$
\beq
G''_n(\omega)=   P_n(\omega)G_0(\omega),
\eeq
where $P_n(\omega)$ is a polynomial that depends on the
$|\Delta_m|^2$'s for $m\leq n$, and is given by
\beq
P_n(\omega)= \prod_{k=1}^{n}|\Delta_k|  \det\left(\omega- L\right)_{n-1}.
\label{eq: pn}
\eeq
In the above, $\left(z-L\right)_{n-1}$ is the upper left $(n-1)
\times (n-1)$ sub-matrix of $(z-L)$.

The determinants $\det\left(z-L\right)_{n-1}$ obey the recursion
relation
\beq
\det\left(z-L\right)_{n+1}=z \det\left(z-L\right)_{n}
+|\Delta_n|^2  \det\left(z-L\right)_{n-1}.
\label{eq: orthogonal poly}
\eeq
Equation~(\ref{eq: orthogonal poly}) is a recursion relation for
orthogonal polynomials. The polynomials $P_n(\omega)$ defined in
Eq.~(\ref{eq: pn}) are therefore orthogonal polynomials under the
scalar product defined by
\beq
\int_{-\infty}^{\infty}d\omega G_0(\omega)
P_n(\omega)P_m(\omega) = \delta_{nm} .
\eeq

The complexity of  computation of $C_n(\beta) , P_n(\omega)$
depends on obtaining all the low $\Delta_m, m\le n$. Therefore,
the expansion (\ref{G-expand2}) would be useful if it could be
truncated at finite $n$, provided that the coefficients
$C_n(\beta)$ decay rapidly  enough with $n$.

\section{Linear recurrents and the Gaussian spectral density}
\label{app: gaussian}
Let us consider a sequence of recurrents $|\Delta_n|^2$ given by
\beq
|\Delta_n|^2=\half n \Omega^2 \qquad n=1,2,\ldots
\label{eq: recurrents linear}
\eeq
The continued fraction representation can be solved using
the following map. Consider the dimensionless position operator $\hx$
represented by raising and lowering operators of the one dimensional
harmonic oscillator
\beq
x=\frac{1}{\sqrt{2}}(a^{\dag}+ a).
\eeq

Since we have $\6 n+1| \hx |n\9=\sqrt{(n+1)/2}$, the function
$G_n(z)$ of Eq.~(\ref{G-expand}) is equivalent to
\beq
G_n(z)=\mu_0\6 n|\frac{1}{z-\Omega\hx}| 0\9.
\eeq
Using the $x$ representation for the ground state of the harmonic
oscillator we have
\bea
G_0(\omega)/\mu_0&=&-\int_{-\infty}^{\infty}dx
\frac{1}{\omega+i\epsilon-\Omega x}\frac{e^{-x^2}}{\sqrt{\pi}}\nonumber\\
&=&i\sqrt{\frac{\pi}{\Omega}}e^{-\omega^2/\Omega^2}\Big(1+\frac{2i}{\pi}\int_0^{\omega^2/\Omega^2}e^{t^2}dt\Big).\nonumber\\
\eea
Likewise, for higher values of $n$ we have
\bea
\im G_n(\omega)/\mu_0&=&- \im \int_{-\infty}^{\infty}dx
\frac{N_n H_n(x)}{\omega+i\epsilon-\Omega x}\frac{e^{-x^2}}{\sqrt{\pi}}\nonumber\\
\nonumber\\
&=&\frac{\sqrt{\pi}}{2}e^{-\omega^2/\Omega^2}N_n
H_n(\omega/\Omega),\nonumber\\
\eea
where $N_n=1/\sqrt{2^n n!}$.

\end{document}